\documentclass[twocolumn,preprintnumbers,amsmath,amssymb,superscriptaddress,10pt]{revtex4}
\usepackage{amssymb}
\usepackage{graphicx,color}
\usepackage{bm}

\begin{document}

\title{Thermalization of Gluons with Bose-Einstein Condensation}

\author{Zhe Xu}
\affiliation{Department of Physics, Tsinghua University and Collaborative Innovation Center of Quantum Matter, Beijing 100084, China}

\author{Kai Zhou}
\affiliation{Department of Physics, Tsinghua University and Collaborative Innovation Center of Quantum Matter, Beijing 100084, China}
\affiliation{Institut f$\ddot{u}$r Theoretische Physik, Johann Wolfgang Goethe-Universit$\ddot{a}$t Frankfurt, Max-von-Laue-Strasse 1, 60438 Frankfurt am Main, Germany}

\author{Pengfei Zhuang}
\affiliation{Department of Physics, Tsinghua University and Collaborative Innovation Center of Quantum Matter, Beijing 100084, China}

\author{Carsten Greiner}
\affiliation{Institut f$\ddot{u}$r Theoretische Physik, Johann Wolfgang Goethe-Universit$\ddot{a}$t Frankfurt, Max-von-Laue-Strasse 1, 60438 Frankfurt am Main, Germany}

\begin{abstract}
We study the thermalization of gluons far from thermal equilibrium in 
relativistic kinetic theory. The initial distribution of gluons is assumed
to resemble that in the early stage of ultrarelativistic heavy ion collisions.
Only elastic scatterings in static, nonexpanding gluonic matter are
considered. At first we show that the occurrence of condensation in the limit
of vanishing particle mass requires a general constraint for the scattering 
matrix element. Then the thermalization of gluons with Bose-Einstein 
condensation is demonstrated in a transport calculation. We see a continuously
increasing overpopulation of low energy gluons, followed by a decrease to 
the equilibrium distribution, when the condensation occurs. The times
of the completion of the gluon condensation and of the entropy production are
calculated. These times scale inversely with the energy density.
\end{abstract}

\maketitle

A new state of matter composed of quarks and gluons, the quark-gluon plasma
(QGP), has been produced in experiments of ultrarelativistic heavy ion
collisions at the Relativistic Heavy Ion Collider (RHIC) and at the Large
Hadron Collider (LHC) \cite{Adams:2005dq,Adcox:2004mh,Aamodt:2010pa}.
The shear viscosity over the entropy density ($\eta/s$) of the QGP,
extracted from the flow measurements, is a small number
\cite{Schenke:2011tv}, which indicates that the QGP is an almost perfect
fluid. However, the initially produced quarks and gluons are
far from thermal equilibrium. How these partons thermalize toward QGP within
a short time of about 1 fm/c is still an open question, although different
approaches \cite{Huang:2014iwa,Xu:2004mz} have been developed to study
this issue.

Recently, a new idea of the formation of Bose-Einstein condensates
in ultrarelativistic heavy ion collisions \cite{Blaizot:2011xf} raises 
again the interest in thermalization of quarks and gluons produced in such
reactions. The condition for a potential condensation is 
an overoccupied initial distribution of gluons, when
compared with the thermal Bose-Einstein distribution with a temperature
obtained by assuming a sudden equilibration with energy and gluon number
conservation. Quantitative studies 
\cite{Blaizot:2013lga,Scardina:2014gxa} have been performed within
the kinetic transport theory, where an overoccupied
initial condition is given by the color glass condensate
formed in high energy heavy ion collisions.
All these works were only able to address the evolution of the gluon
system till the onset of Bose-Einstein condensation.
Whether or not a gluon condensate appears
from kinetic statistical studies is an important question.
In this Letter we evolve the gluon system beyond the onset toward
the full thermalization with a complete Bose-Einstein condensation.
It is an important step forward toward understanding the parton thermalization
in ultrarelativistic heavy ion experiments.
Our work demonstrates an example of Bose-Einstein condensation and
thermalization in a relativistic system being far from thermal
equilibrium and with high energy density, which also provides new
insights for other field of physics.

In this work, the space-time evolution of gluons with the Bose-Einstein
condensation is described in a kinetic transport approach 
BAMPS (Boltzmann approach of multiparton scatterings) \cite{Xu:2004mz}
that solves the Boltzmann equation for gluons.
We consider a static, nonexpanding gluon system that is initially
out of equilibrium and assume gluon number conservation by
including only gluonic elastic scatterings in the collision term of
the Boltzmann equation.
Once the initial distribution of gluons is given, the gluon system
will finally evolve to the thermal equilibrium distribution
\begin{equation}
\label{equil}
f_{eq}({\mathbf p})=\frac{1}{e^{E/T_{BE}}-1}+(2\pi)^3 n_c^{eq} \delta^{(3)}({\mathbf p}) \,,
\end{equation}
which is the solution of the Boltzmann equation in the long time limit.
The temperature $T_{BE}$ and the density of the gluon condensate
at equilibrium $n_c^{eq}$
are obtained by using the assumed gluon number and energy conservation.
We will show that our numerical solution agrees with Eq. (\ref{equil}),
and will determine the time scale of the full thermalization for various
initial conditions from weak to strong overoccupation.

We note that although particle number changing processes may prevent
the Bose-Einstein condensation, studies within field theories including
particle production and annihilation processes showed the emergence
of a condensate for an intermediate time window \cite{Epelbaum:2011pc}.
Also, results from kinetic transport calculations indicate an acceleration
of the onset of gluon condensation due to gluon bremsstrahlung 
processes \cite{Huang:2013lia}.

The formation of a Bose-Einstein condensate is the consequence of
quantum statistics of bosons and has been observed in experiments using 
ultracold atoms \cite{Anderson:1995gf}.
The Boltzmann equation including Bose statistics reads
\begin{eqnarray}
\label{boltzmann}
&&\left ( \frac{\partial}{\partial t} + \frac{{\mathbf p}_1}{E_1}
\frac{\partial}{\partial {\mathbf r}} \right )\, 
f_1 = \frac{1}{2E_1} \int d\Gamma_2 \frac{1}{2} \int d\Gamma_3 d\Gamma_4
| {\cal M}_{34\to 12} |^2 \nonumber \\
&& \times \ \left[ f_3 f_4 (1+f_1) (1+f_2) - f_1 f_2 (1+f_3) (1+f_4) \right] 
(2\pi)^4 \nonumber \\
&& \times \ \delta^{(4)} (p_3+p_4-p_1-p_2) \,,
\end{eqnarray}
where $f_i=f_i({\mathbf r}, {\mathbf p}_i, t)$ and 
$d\Gamma_i=d^3p_i/(2E_i)/(2\pi)^3, i=1,2,3,4$. Binary collisions 
$34\to12$ and $12\to34$ are determined by the collision kernel
$| {\cal M}_{34\to 12} |^2$ and  $| {\cal M}_{12\to 34} |^2$, 
which are equal. 
$(1+f_1)(1+f_2)$ and $(1+f_3)(1+f_4)$ are the Bose factors, with which
the distribution (\ref{equil}) is a solution of Eq. (\ref{boltzmann}).

We decompose the distribution function $f$ into two parts
$f=f^{gas}+f^{cond}$, where $f^{gas}$ denotes the distribution of gas
(noncondensate) particles, while 
$f^{cond}=(2\pi)^3 n_c \delta^{(3)}({\mathbf p})$ denotes the distribution
of the condensate particles with zero momentum. $n_c$ is the density of
the condensate particles, which grows with time during the condensation.
We obtain easily the equation describing the condensation of zero momentum
particles, when we replace $f_1$ and $1+f_1$ in Eq. (\ref{boltzmann}) by
$f^{cond}_1({\mathbf r}, {\mathbf p}_1, t)$. Integrating this equation over
${\mathbf p}_1$ gives the time derivative of $n_c$.
After a lengthy integration with the help of the delta functions 
$ \delta^{(3)}({\mathbf p}) \delta^{(4)} (p_3+p_4-p_1-p_2)$ we have 
(as that derived in \cite{Semikoz:1994zp})
\begin{eqnarray}
\label{condrate}
\frac{\partial n_c}{\partial t} = \frac{n_c}{64\pi^3} 
&\int& dE_3 dE_4 \left [ f_3 f_4 - f_2 (1+f_3+f_4) \right ] \nonumber \\
&& \times \ E \left[ \frac{| {\cal M}_{34\to 12} |^2}{s} \right]_{s=2mE} \,.
\end{eqnarray}
The two terms on the right-hand side correspond to kinetic processes for
the condensation and the evaporation, respectively.
In the derivation we have assumed that $f$ only depends on the absolute
value of ${\mathbf p}$. The term of the spacial derivative of $n_c$ 
drops out, since we consider a homogeneous gluon matter.
$m$ denotes the particle mass at rest. $E=E_3+E_4$ is the total
energy in the collision, while $p=|{\mathbf p}_3+{\mathbf p}_4|$
is the total momentum. $s=E^2-p^2$ is the invariant mass.
The energy-momentum conservation leads to $E=m+E_2$ and $p=p_2$, where
particle $1$ denotes the condensate particle with zero momentum.
From $E_2^2=p_2^2+m^2$ we obtain $s=2mE$, which is the kinematic
constraint for condensation processes. 

Equation (\ref{condrate}) indicates a general condition for the
occurrence of the Bose-Einstein
condensation: $| {\cal M} |^2/s$ must be finite. In particular,
this condition should hold for vanishing rest mass of particles $m=0$,
which is the case we are considering in this work. The elastic
scatterings of massless gluons are described in leading order of
perturbative QCD. We use
\begin{equation}
\label{matrix}
| {\cal M}_{gg\to gg} |^2 \approx 144 \pi^2 \alpha_s^2 \frac{s^2}{t (t-m^2_D)}
\end{equation}
as the collision kernel, which has been calculated by using the standard 
Hard-Thermal-Loop (HTL)
treatment according to a sum rule satisfied by the HTL gluon propagator  
\cite{Aurenche:2002pd,Kurkela:2011ti}. $m_D$ is the Debye screening mass.
In a process $34\to 12$,
where ``1'' denotes a condensate gluon with zero momentum and zero mass,
gluon $2$, $3$ and $4$ must be colinear. Thus, the Mandelstam variables $s$ and $t$
are zero and $s/t$ goes to $-1$. We see that 
$[| {\cal M}_{gg\to gg} |^2/s]_{s=0}\sim \alpha_s^2/m_D^2$ is finite.
On the contrary, the often used kernel \cite{Wong:1996va}
$| \tilde {\cal M} |^2 \sim s^2/(t-m^2_D)^2$ gives
$[| \tilde {\cal M} |^2/s]_{s=0}=0$, which indicates that the gluon
condensation will not happen, although the gluon system has evolved to
the onset.

We note that for scatterings involving noncondensate particles only,
i.e., $s >0$, the total cross section $\sigma_{gg\to gg}$ is logarithmically
divergent due to Eq. (\ref{matrix}). This is regularized by an upper cutoff of $t$,
since scatterings with $t$ approaching to zero do not contribute to thermalization.

We have derived Eq. (\ref{condrate}) in order to prove the justification
of the chosen collision kernel Eq. (\ref{matrix}). The time evolution of the
condensation and thermalization is achieved by solving the Boltzmann equation
(\ref{boltzmann}) numerically. To do it we modify the currently used partonic transport
model BAMPS \cite{Xu:2004mz}
to include the Bose factors. 
A new scheme \cite{jackson} is employed to simulate collisions stochastically
between test particles. Instead of a collision probability we define
a differential collision probability
\begin{equation}
\label{dp22}
\frac{dP_{22}}{d\Omega}=\frac{v_{rel}}{N_{test}} \frac{d\sigma_{22}}{d\Omega}
(1+f_3) (1+f_4) \frac{\Delta t}{\Delta V} \,.
\end{equation}
$v_{rel}=s/(2E_1E_2)$ is the relative velocity of the incoming particles,
$N_{test}$ the number of test particles per real particle, $\Delta V$
the volume element around the collision point, $\Delta t$ the time step
in the calculation, and $d\sigma_{gg\to gg}/d\Omega$ the differential
cross section of elastic gluonic scatterings. 
At first a solid angle $\tilde \Omega$ is sampled according to 
$d\sigma_{22}/d\Omega$ for each of the particle pairs considered. In case of
the occurrence of a collision the momenta ${\mathbf p}_3, {\mathbf p}_4$
of the outgoing particles are thus determined. Kinematic constraints
ensure exact energy-momentum conservation in each of the collisions.
We also obtain $f_3$ and $f_4$
from the extracted $f$ at ${\mathbf p}_3$ and ${\mathbf p}_4$, respectively.
Second, a random number between zero and the value of a normalized reference
function $dF/d\Omega$ at $\tilde \Omega$ will decide whether a collision
will occur: if this random number is smaller than the differential collision
probability $dP_{22}/d\Omega$ at $\tilde \Omega$, a collision occurs;
otherwise not.
In processes $12\to 3c$ where a condensate gluon $c$ is produced,
the two incoming gluons should have parallel momenta, which is
impossible in numerical calculations.
We define a cutoff energy $\epsilon$. Gluons with energy smaller
than $\epsilon$ are regarded as condensate gluons. 
Since the production rate of the condensate particles should not depend 
on the particular numerical implementation, $\epsilon$ has to be chosen
sufficiently small to avoid numerical artifacts.
The details and numerical confirmation of the new scheme will be
shown in a forthcoming paper \cite{zhou}.

The momentum distribution of initial gluons is assumed to be
\begin{equation}
\label{cgc}
f_{init}({\mathbf p})=f_0 \theta (Q_s-|{\mathbf p}|) \,,
\end{equation}
which resembles that in the early stage of ultrarelativistic heavy ion
collisions \cite{Blaizot:2011xf}.
The initial momentum distribution has been
simplified to be isotropic, although the new method introduced \cite{zhou}
can be applied for anisotropic momentum distributions.
From (\ref{cgc}) we obtain the temperature at equilibrium and the
condensate fraction of the total density
\begin{equation}
\label{temp}
T_{BE}=\left ( 15f_0/4 \right )^{1/4} Q_s/\pi\,, \
n_c^{eq}/n=1-19.43\pi^{-3}f_0^{-1/4} \,.
\end{equation}
There is a critical value of $f_0^c=0.154$, below which $n_c^{eq}/n$
becomes negative and no Bose-Einstein condensation occurs.
We set $Q_s=1$ GeV and vary $f_0$ from 0.4 to 2.0 to change the number and
energy density. Gluons are homogeneously distributed in a box, for which
the side length is set to be 3 fm. A periodic boundary condition is taken.
The box is divided by cubic cells with equal volume $\Delta V$. The side
length of a cell is 0.125 fm. Large values of $N_{test}$ are used to enhance
statistics and to reduce numerical uncertainties in the extraction of $f$.
For scatterings a constant coupling $\alpha_s=0.3$ is used and the screening
mass is evaluated at each of the time steps by \cite{Wong:1996va}
\begin{equation}
\label{md2}
m^2_D=16\pi N_c \alpha_s \int \frac{d^3p}{(2\pi)^3} \, \frac{1}{p} f^{gas}\,.
\end{equation}

\begin{figure}[h]
 \centering
 \includegraphics[width=0.45\textwidth]{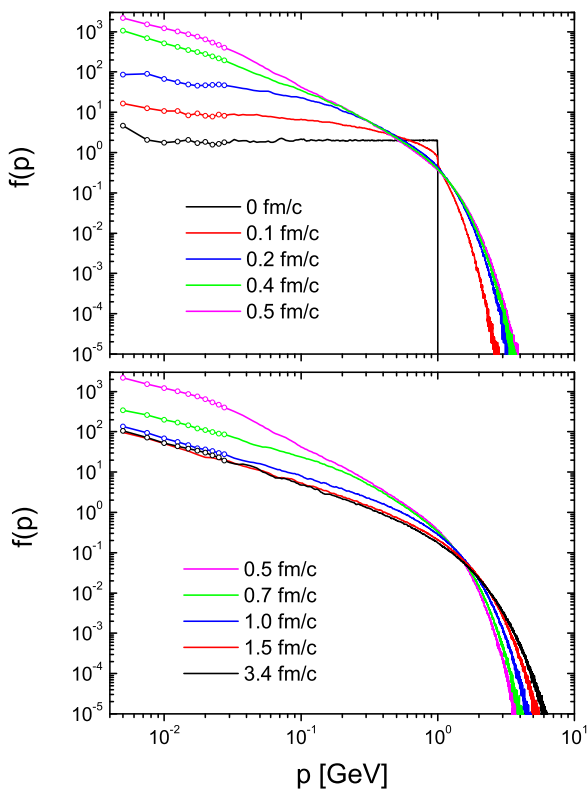}
 \caption{(color online) The energy distributions of noncondensate gluons
at various times. $f_0=2$ is chosen.}
 \label{spec}
\end{figure}

\begin{figure}[h]
 \centering
  \includegraphics[width=0.45\textwidth]{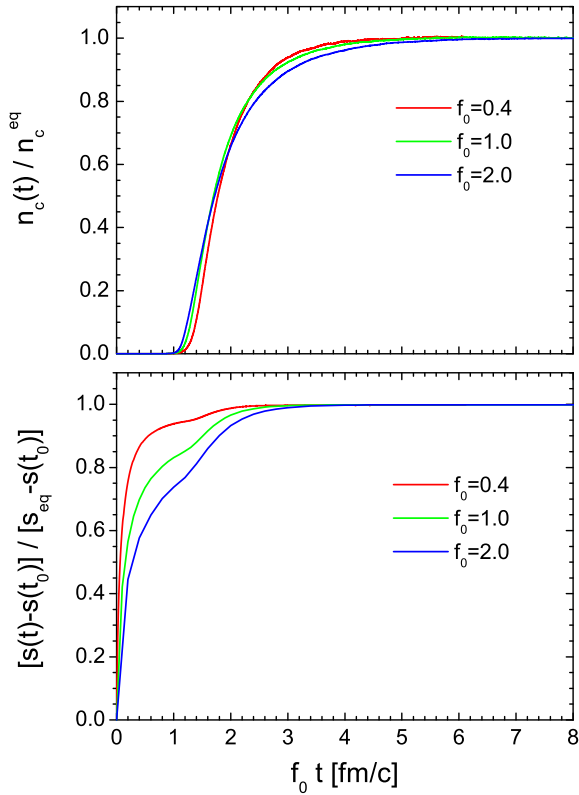}
 \caption{(color online) Time evolution of the density of the gluon
 condensate (upper panel) and the entropy production (lower panel).
 Calculations are performed for various $f_0$.}
 \label{bec}
\end{figure}

Figure \ref{spec} shows the energy distributions of noncondensate gluons
at various times.
$f_0=2.0$ is chosen in the example. From the upper panel of Fig. \ref{spec}
we observe a continuous increase at low energy and at energy higher than $Q_s$,
whereas between some intermediate energy and $Q_s$ the distribution
decreases continuously. This indicates that particle and energy are flowing
from the scale at $Q_s$ toward both lower and higher energies at early times
of the thermalization process.
The distributions shown in Fig. \ref{spec} are calculated by counting 
test particles in equal energy intervals of $\Delta p=0.005$ GeV
and dividing these particle numbers by $N_{test}Vp^2\Delta p/(2\pi^2)$,
where $V$ is the volume of the box and $p$ is the middle position
of the energy intervals. The distributions are plotted at each momenta
$p$. The open symbols in Fig. \ref{spec} are first ten values of these
distributions. The particle number in intervals at low energies is smaller
and thus has a larger fluctuation compared to those at intermediate energies.
For the present setups it is difficult to make an unambiguous parametrization
for the distributions at low energies. One needs a larger value of $N_{test}$
for obtaining higher accuracy. Nevertheless, we fit the distributions at 
low energies
by Bose-Einstein distributions with different effective temperatures $T_{eff}$
and chemical potentials $\mu_{eff}$. We find that $\mu_{eff}$ increases
from a negative value to zero, which indicates the onset
of the Bose-Einstein condensation \cite{Blaizot:2013lga,Scardina:2014gxa}.
When $\mu_{eff}$ is close to zero, its value will fluctuate around zero
due to the numerical uncertainty. Once $\mu_{eff}$ becomes positive, we
define this moment as the start time of the gluon condensation $\tau_c$. 
At this start time $n_c$ probably has a nonzero value, because we have
made the approximation to regard particles with energy smaller than 
$\epsilon$ as condensate gluons. In the calculations we set 
$\epsilon=0.0025$ GeV, which is small enough to reduce numerical
errors. We see later in Fig. \ref{bec} that $n_c$ at long time 
agrees perfectly with the analytical equilibrium value $n_c^{eq}$.

We note that both $\tau_c$ and $n_c$ at $\tau_c$ should be determined
from a theory describing the phase transition to a Bose-Einstein 
condensate \cite{Epelbaum:2011pc}, which is beyond the scope of the
Boltzmann approach. However, we will see these values will not
affect the dynamic behavior of the growth of the condensate appreciably.
In the calculation for $f_0=2.0$ the condensation starts at 
$\tau_c=0.376$ fm/c. The energy distribution at $\tau_c$
is far from thermal equilibrium. This particular feature differs
from the production of Bose-Einstein condensates in experiments using 
ultracold atoms \cite{Anderson:1995gf}, where the system is close to thermal
equilibrium. From the upper panel of Fig. \ref{spec} we also see
that until $0.5$ fm/c the condensation is too weak to reduce the increasing
overpopulation at low energies. $T_{eff}$ increases to $10$ GeV at $0.5$ fm/c,
which is much larger than the equilibrium temperature, $T_{BE}=0.53$ GeV.

After $0.5$ fm/c the overpopulation at low energies begins to disappear
due to the growth of the gluon condensate, as clearly seen in the lower
panel of Fig. \ref{spec}. We find that the distribution at low energies
has a profile $f \sim 1/p$ from $\tau_c$.
The freed energy during the condensation flows
to particles with high energies, where the occupation number increases
continuously. The energy distribution relaxes to the equilibrium form.
We find a perfect agreement with the equilibrium distribution at 3.4 fm/c,
which indicates the completion of the gluon condensation, see Fig. \ref{bec}.
We notice that the distribution at 1.5 fm/c is already very close to
the equilibrium one.

The growth of the gluon condensate in time is shown in the upper panel
of Fig. \ref{bec}.
At early times of the condensation the gluon condensate grows exponentially,
which can qualitatively be understood by Eq. (\ref{condrate}).
At these times the production processes are dominant. When the evaporation
processes begin to balance the production processes, the growth of the
gluon condensate slows down, and then has a relaxation form. 

Numerical calculations are performed for three different $f_0$. Larger $f_0$
leads to larger particle and energy densities and larger fraction of the
density of the gluon condensate to the total gluon density, see
Eq. (\ref{temp}). To study the scaling behavior, the density of the gluon
condensate is divided by its final value at equilibrium, while the time is
multiplied by $f_0$. We see an approximate scaling, which shows that
{\em the larger the density (larger $f_0$), the faster is the completion
of the gluon condensation, and the faster is the thermal equilibration}.
The scaling behavior of the starting time of the gluon condensation
$\tau_c$ can be understood by approximating the collision term in the
Boltzmann equation (\ref{boltzmann}),
$f_3f_4(1+f_1)(1+f_2)-f_1f_2(1+f_3)(1+f_4)\approx f_3f_4(f_1+f_2)-f_1f_2(f_3+f_4)$.
It is a good approximation for colliding particles with small energy,
where $f$ is large. If $f$ at low energies is proportional to $f_0$,
$f \sim f_0 f_N(p,t)$, which seems reasonable at least at early times due
to the given initial condition (\ref{cgc}), and if this relation holds long
for large $f_0$, then the right-hand side of (\ref{boltzmann}) scales
with $f_0^2$ by noting that $| {\cal M}_{gg\to gg} |^2\sim 1/m_D^2$ and
$m_D^2$ is proportional to $f_0$. The Boltzmann equation (\ref{boltzmann})
can be reduced to an equation for $f_N(p,t')$ with $t'=f_0t$.
Since $f_N(p,t')\approx T_{eff}(t')/f_0/[p-\mu_{eff}(t')]$ for low energy,
$\mu_{eff}$ depends on $t'$, which explains the scaling behavior of $\tau_c$,
when $\mu_{eff}$ approaches zero with $t'$. In addition, $T_{eff}$ is
proportional to $f_0$ at $\tau_c$, which is also confirmed from our numerical
results. After $\tau_c$ the scaling of $n_c$ can be understood analogously
by using $f_3f_4-f_2(1+f_3+f_4)\approx f_3f_4-f_2(f_3+f_4)$ in 
Eq. (\ref{condrate}), if the relation $f\sim f_0f_N$ still holds.
At late times when $n_c$ relaxes to its final value, $f$ relaxes to the
Bose-Einstein distribution and is not proportional to $f_0$ even at low
energies. This violates slightly the time
scaling of $n_c$, which is seen in Fig. \ref{bec} at late times.
On the contrary, the differences of $n_c(t')$ shortly after $\tau_c$
are due to numerical uncertainties in $\tau_c$ as well as $n_c$ at $\tau_c$.

The lower panel of Fig. \ref{bec} shows the entropy production
$s(t)-s(t_0=0)$ divided by its value at equilibrium.
We see a two step production separated by $\tau_c$. This shows that
the formation of the gluon condensate is essential for the time
scale of thermalization. From Fig. \ref{bec} we also see that the
completion of the condensation is later than that of entropy production.
The full thermalization is reached at $f_0 t\approx 3$ fm/c.
In addition, the entropy production does not scale with $f_0t$ before
$\tau_c$, because the momentum distribution at low energies has a negligible
contribution to the entropy and the argument for $\tau_c$ scaling does
not apply to the early entropy production. Instead, we find that
the early entropy production is independent of $f_0$.
The reason lies in the almost same collision rate at the early
times, because $\sigma_{gg\to gg}\sim 1/m_D^2$ and $m_D^2$ is
roughly proportional to $f_0$.

In this Letter we demonstrated for the first time a complete Bose-Einstein
condensation of gluons within kinetic theory. Our calculations
showed that the times for
the occurrence of the condensation, the completion of the condensation, and
the full thermalization scale inversely with the energy density. 
For instance, for energy density $53 \mbox{ GeV/fm}^3$ ($f_0=2$) the time of
the full thermalization is about $1.5$ fm/c, which is consistent with
the parametric result $t_{th} \sim Q_s^{-1} \alpha_s^{-7/4}$
found in \cite{Blaizot:2011xf}. We also
find that the earlier the condensation occurs, the earlier does the second
step of the entropy production start. All these are important new findings
to get further forward with the understanding of thermalization for systems
far from thermal equilibrium.

As discussed in \cite{Voskresensky:1994uz,Chiu:2012ij,McLerran:2014hza},
pions, photons, and dileptons produced in ultrarelativistic heavy ion collisions
might provide measurable hints for the formation of a gluon condensate.
The experimental significance also depends on how long the gluon
condensate lives after its formation.
In our present study a gluon condensate exits, because we assumed the
gluon number conservation and considered a static, nonexpanding system.
The inclusion of quark-antiquark annihilations \cite{Blaizot:2014jna},
or/and gluon bremsstrahlung processes, and the back reactions
 \cite{Huang:2013lia,York:2014wja} will violate the gluon number
conservation, which will reduce the overoccupation of gluons at small momenta.
Whether a gluon condensation will still occur, is an interesting question.
If the condensation in fact happens, the gluon condensate is only an intermediate state
and may disappear in the later evolution.
Moreover, one has to consider the expansion of quark gluon matter, in order to
obtain more realistic results.  All these are subjects of future works.

Z.X. thanks X.G. Huang, J. Liao, V. Greco, and L. McLerran for helpful discussions.
This work was financially supported by the MOST, the NSFC under Grants No. 2014CB845400,
No. 11275103, and No. 11335005. The BAMPS simulations were performed 
at Tsinghua National Laboratory for Information Science and Technology.

\end{document}